\documentclass[]{raa}            
\usepackage{graphicx,times}
\usepackage{natbib}

\providecommand{\bm}[1]{\mbox{\boldmath$#1$\unboldmath}}

\begin{document}

   \title{Eight new quasars discovered by LAMOST in one extragalactic field 
}

 \volnopage{ {\bf 2010} Vol.\ {\bf X} No. {\bf XX}, 000--000}
   \setcounter{page}{1}

   \author{Xue-Bing Wu\inst{1}, Zhendong Jia\inst{1}, Zhaoyu Chen\inst{1}, Wenwen Zuo
      \inst{1}
   \and Yongheng Zhao\inst{2}, Ali Luo\inst{2}, Zhongrui Bai\inst{2}, Jianjun Chen\inst{2}, Haotong Zhang\inst{2}, Hongliang Yan\inst{2}, Juanjuan Ren\inst{2}, Shiwei Sun\inst{2}, Hong Wu\inst{2},
   \and Yong Zhang \inst{3}, Yeping Li \inst{3}, Qishuai Lu \inst{3}, You Wang \inst{3}, Jijun Ni \inst{3}, Hai Wang \inst{3}, Xu Kong\inst{4}, Shiyin Shen\inst{5}
   }

   \institute{ Department of Astronomy, Peking University, Beijing 100871, 
China; {\it wuxb@bac.pku.edu.cn}\\
        \and
             National Astronomical Observatories, Chinese Academy of Sciences,
             Beijing 100012, China\\
\and 
National Institute of Astronomical Optics \& Technology, Chinese Academy of
Science, Nanjing 210042, China
\and 
Center for Astrophysis, University of Science \& Technology of China, Hefei 230026, China
\and
Shanghai Astronomical Observatory,  Chinese Academy of Sciences, Shanghai 200030, China
\\
\vs \no
   {\small Received [year] [month] [day]; accepted [year] [month] [day] }
}

\abstract{ We report the discovery of eight new quasars in one extragalactic 
field (five degree centered at RA=$08^h58^m08.2^s$, Dec=$01^o32'29.7''$) with the 
LAMOST commissioning observations on December 18, 2009. 
These quasars, with $i$ magnitudes from 16.44 to 19.34 and redshifts from 0.898
to 2.773,  were not identified in the SDSS spectroscopic survey, though six of 
them with redshifts less than 2.5 were selected as quasar targets in SDSS.
Except one source without near-IR $Y$-band data, seven of these eight new quasars
meet a newly proposed quasar selection
criterion involving both near-IR and optical colors. Two of them were found
in the 'redshift desert' for quasars ($z$ from 2.2 to 3) , indicating that the new criterion
is efficient for recovering the missing quasars with similar optical colors as
stars. Although LAMOST met some problems 
during the commissioning 
observations, we were still able to identify other 38 known SDSS 
quasars in this field, with $i$ magnitudes from 16.24 to 19.10 and redshifts 
from 0.297 to 4.512. Our identifications imply that a substantial fraction of quasars
may be missing in the previous quasar surveys. The implication of our results 
to the future LAMOST quasar survey is discussed.
\keywords{quasars: general --- quasars: emission lines --- galaxies: active
}
}

   \authorrunning{X.-B. Wu et al. }            
   \titlerunning{Eight new quasar discovered by LAMOST}  
   \maketitle


%
%
\section{Introduction}           
\label{sect:intro}
Quasars are interesting objects in the universe since they can be used as 
important tools to probe
the accretion power around supermassive black holes, the intergalactic 
medium, the large scale structure and the cosmic reionization.
The number of quasars has increased
steadily in the past four decades (Richards et al. 2009). Especially, 
A large number of them have 
been discovered in two spectroscopic surveys, namely, the Two-Degree 
Fields (2dF) survey (Boyle et al. 2000) and Sloan Digital Sky Survey (SDSS)
(York et al. 2000). 2dF has discovered more than 20,000 low redshift ($z<2.2$) 
quasars with UV-excess (Croom et al. 2004, Smith et al. 2005),
while SDSS has identified more
than 100,000 quasars (Schneider et al. 2010; Abazajian et al. 2009). 
Some dedicated methods were proposed for finding higher redshift quasars 
(Fan et al. 2001a,b; Richards et al. 2002). However, the efficiency
of identifying quasars with redshift between 2.2 and 3 is still very low in SDSS
(Schneider et al. 2010). This is mainly because quasars with such redshifts 
have very similar optical 
colors as stars and are mostly excluded by the SDSS spectroscopy. Therefore, 
the redshift range
from 2.2 to 3 is regarded as the `redshift desert' for quasars because
of the difficulty in identifying quasars with redshifts in this range. 

However, quasars in the redshift desert are usually more luminous than normal 
stars in the infrared K-band (Warren et al. 2000) becasue the
spectral energy distributions (SEDs) of quasars are flat. 
This provides us an 
important way of finding these quasars by involving the near-IR colors.
Some methods have been suggested by using the infrared K-band excess based on 
the UKIRT
(UK Infrared Telescope) Infrared Deep Sky Survey (UKIDSS) (Warren et al. 2000;
Hewett et al. 2006; Maddox et al. 2008). Combining the UKIDSS YJHK and SDSS
ugriz magnitudes, some criteria to select quasars have been
proposed (Maddox et al. 2008; Chiu et al. 2007).  More recently, based on a large
SDSS-UKIDSS quasar sample, Wu \& Jia (2010) 
proposed to use the $Y-K$ vs. $g-z$ diagram to select $z<4$ quasars and use 
the $J-K$ vs.$i-Y$ diagram to select $z<5$ quasars. Although the success of
adopting these criteria has been demonstrated by using the existing quasar 
sample, we still need to apply them to discover new quasars and investigate
how many quasars missed in the previous spectroscopic surveys.
 
The Large Sky Area Multi-Object Fibre Spectroscopic Telescope (LAMOST) 
is a powerful instrument for spectroscopy (Su et al. 1998) and
the main construction was finished in 2008. Since 2009 LAMOST has entered 
its commissioning 
phase, and some test observations have been done in the winter of 2009. 
Although LAMOST has not reached its full capability, these
observations already led to the discovery of new quasars, including
12 quasars behind M31 (Huo et al. 2010) and a very bright $i=16.44$ quasar with
redshift $z=2.427$ (in the redshift desert) (Wu et al.
2010; hereafter paper I). In this paper, we report the discovery of
more quasars in the same extragalactic field where the very bright quasar
was found, including another $z=2.773$ quasar in the redshift desert.


\section{Target selection and Observation}
\label{sect:Obs}

In the winter of 2009, we have selected several extragalactic fields for the 
LAMOST commissioning observations.
In order to test whether the newly proposed quasar selection criterion in the 
$Y-K$ vs. $g-z$ diagram is efficient in identifying quasars (Wu \& Jia 2010), 
we selected
quasar candidates in several sky fields overlapped between UKIDSS and SDSS
surveyed area. Some additional quasar candidates from the catalog of 
Richards et al. (2009) are also included.
Besides these quasar candidates, we also included many known
SDSS quasars in these fields as targets in order to compare the LAMOST 
spectroscopy with SDSS. Here we report the observational results in one
of these fields, which is a five degree field centered at
RA=$08^h58^m08.2^s$, Dec=$01^o32'29.7''$ close to the field of GAMA-09
(Robotham et al. 2010).   

On December 18, LAMOST made the spectroscopic observations on this field and 357
quasar targets mostly with $i<19.1$ together with other objects were observed with 
the exposure time of 30 minutes and the spectral resolution of $R\sim1000$. 
The spectra were processed using a preliminary 
version of LAMOST spectral pipeline. Due to the problems in many aspects 
during the LAMOST commissioning observations, the overall quality of the spectra is not
satisfactory. Only 99 of 357 quasar targets show the obvious spectral features
of quasars or stars/galaxies, and the rest spectra show either too low S/N 
(signal to noise ratio) or sky light emissions only. Among these 99 objects, 46 
of them can
be identified as quasars and 53 of them show the features of either stars or galaxies.
8 of 46 idenified quasars are new and 38 of them are known SDSS quasars. 
Among
8 new quasars, SDSS J085543.40-001517.7 is a very bright one ($i=16.44$) and
was identified as a $z=2.427$ quasar. This is the first quasar in the
redshift desert discovered by LAMOST and its detailed properties have been 
reported in paper I. For the completeness, we also include some of its 
properties in this paper.     

In Fig. 1 we show the SDSS finding charts\footnote{
Obtained from http://cas.sdss.org/dr7/en/tools/chart/chart.asp}
of 8 new quasars in an order of increasing RA. Clearly they
all are point sources in the optical bands. We also checked their morphology
types in the UKIDSS images and all of them are also point sources  
(UKIDSS mergedclass=-1) in the near-IR bands. This is consistent with the 
morphology type of SDSS-UKIDSS quasars with redshifts larger than 0.5 (Wu \& Jia 2010).
In Table 1 we list the main properties of these 8 quasars, including their
coordinates, magnitudes and
redshifts. The SDSS $ugriz$ magnitudes are given in AB systems and UKIDSS
$YJHK$ magnitudes are given in Vega system. All magnitudes are corrected for
the Galactic extinction using the map of Schlegel et al. (1998). The offsets
between the SDSS and UKIDSS positions are less then $0.21''$ for these 8
quasars, indicating that the mis-identifications of their UKIDSS counterparts
of these SDSS sources are very unlikely.

\begin{table}
\bc
\begin{minipage}[]{100mm}
\caption[]{Parameters of eight new quasars}\end{minipage}
\setlength{\tabcolsep}{0.8pt}
\small
 \begin{tabular}{ccccccccccccc}
  \hline\noalign{\smallskip}
Name& RA& Dec&$u$& $g$& $r$& $i$& $z$& $Y$& $J$& $H$& $K$& LAMOST\\
(SDSS J)& ($^{o}$)&($^{o}$)&&&&&&&&&&redshift\\
  \hline\noalign{\smallskip}
085307.31+014523.1&133.28049&1.75643&17.715&17.718&17.716&17.566&17.453&17.018&16.767&16.435&15.873&1.952\\
085543.40-001517.7&133.93086&-0.25493&17.668&16.866&16.617&16.444&16.208&15.573&15.214&14.585&13.834&2.427\\
085718.29+024017.7&134.32625&2.67160&18.520&18.373&18.128&18.194&18.313&---&---&16.985&16.209&1.154\\
085727.85+012802.1&134.36605&1.46728&18.480&18.489&18.162&18.152&18.258&17.484&17.143&16.344&16.015&1.363 \\
090148.15+004225.9&135.45065&0.70722&19.777&19.513&19.358&19.345&19.130&18.255&17.462&17.181&16.388&0.898 \\
090437.02+014055.3&136.15428&1.68203&18.585&18.570&18.356&18.072&18.009&17.488&---&16.542&15.917&1.765\\
090453.24-001426.5&136.22187&-0.24069&19.433&19.261&19.193&18.917&18.889&18.438&18.033&17.390&16.974&1.670\\
090504.87+000800.5&136.27030&0.13348&20.457&18.863&18.440&18.154&18.162&17.446&16.986&16.515&16.022&2.773 \\
  \noalign{\smallskip}\hline
\end{tabular}
\ec
\tablecomments{0.86\textwidth}{The SDSS $ugriz$ magnitudes are given in AB system 
and the UKIDSS $YJHK$ magnitudes are given in Vega system.}
\end{table}

   \begin{figure}
   \centering
   \includegraphics[width=4.7cm, angle=0]{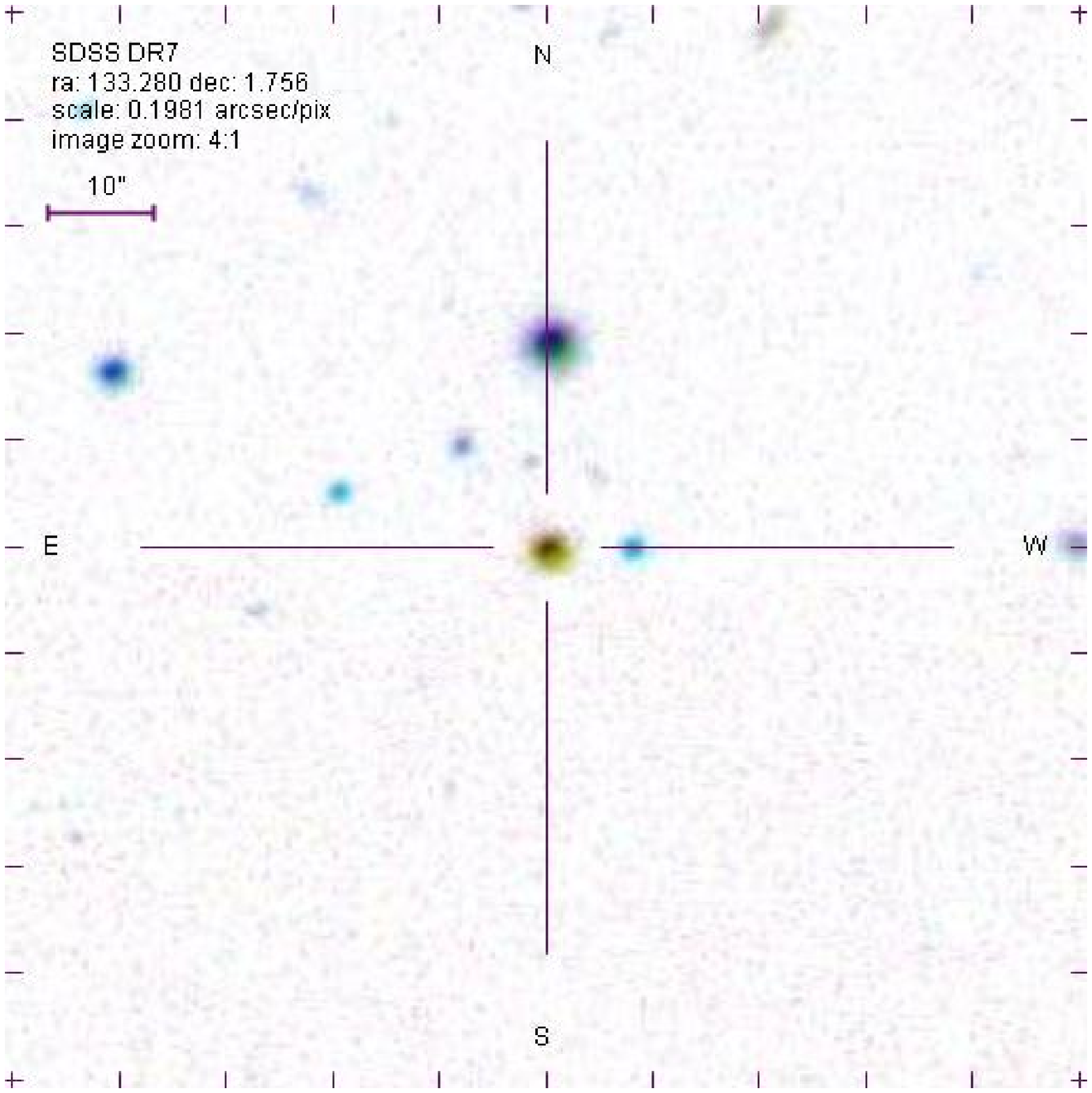}
   \includegraphics[width=4.7cm, angle=0]{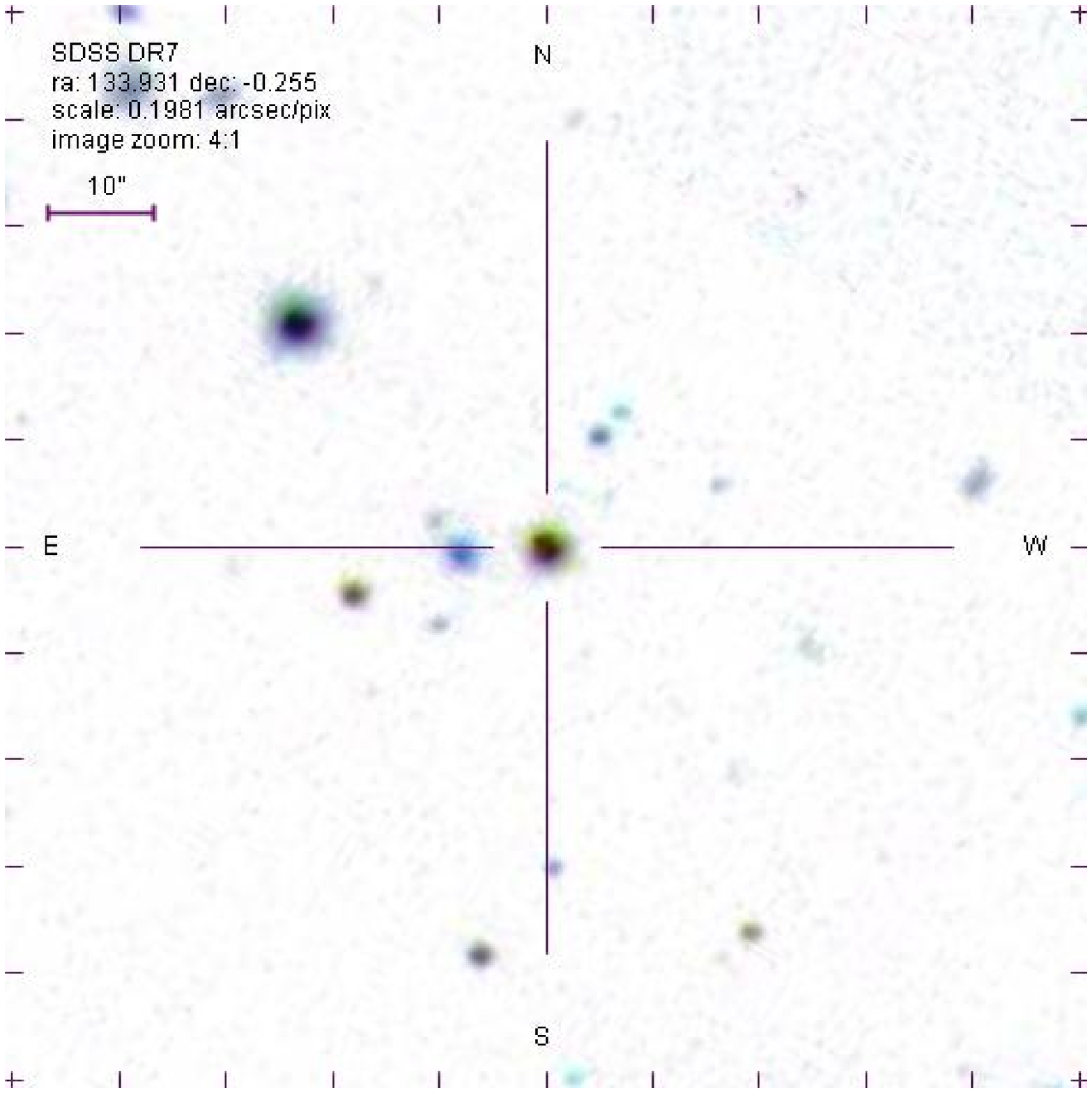}
    \includegraphics[width=4.7cm, angle=0]{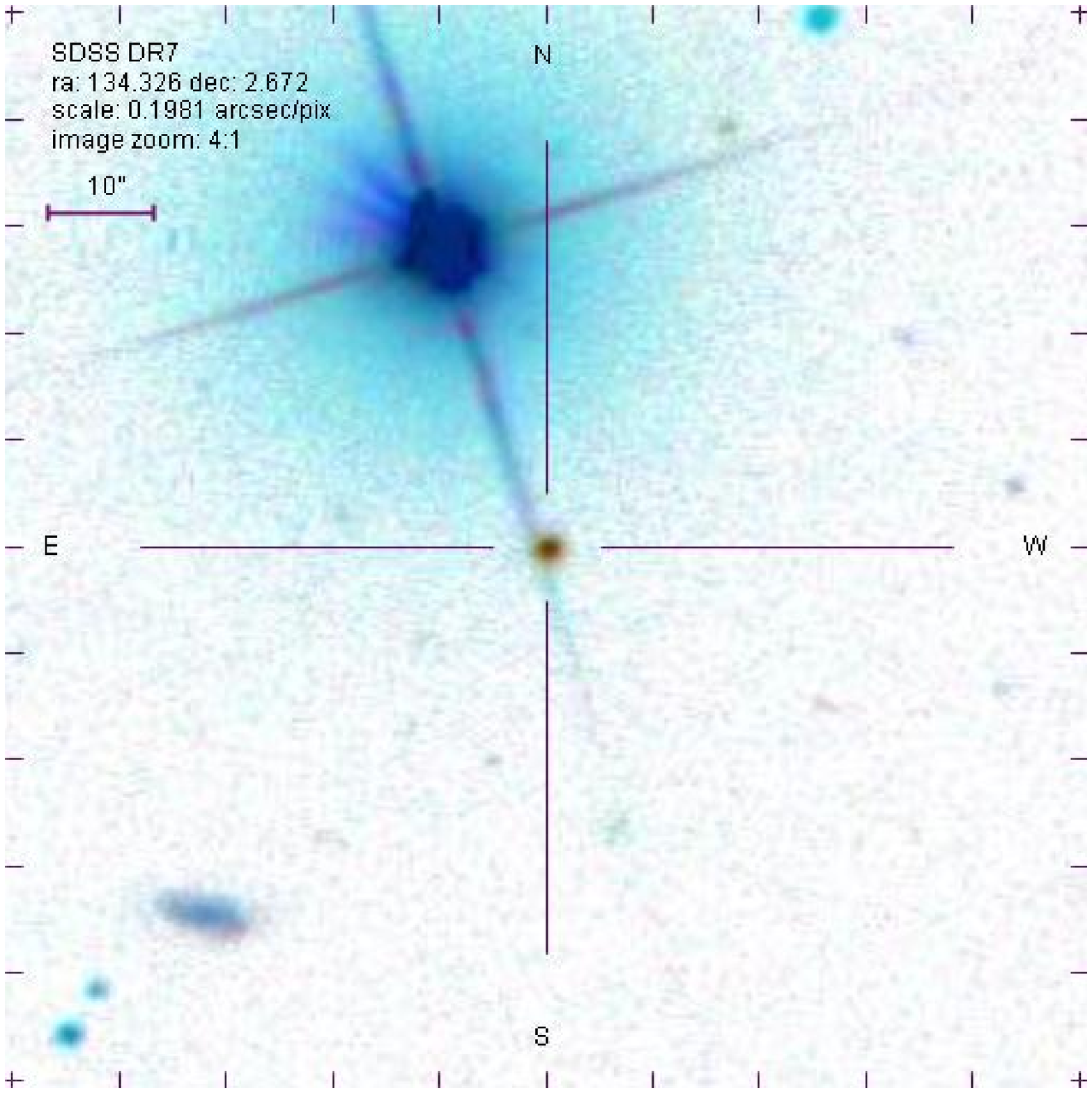}
    \includegraphics[width=4.7cm, angle=0]{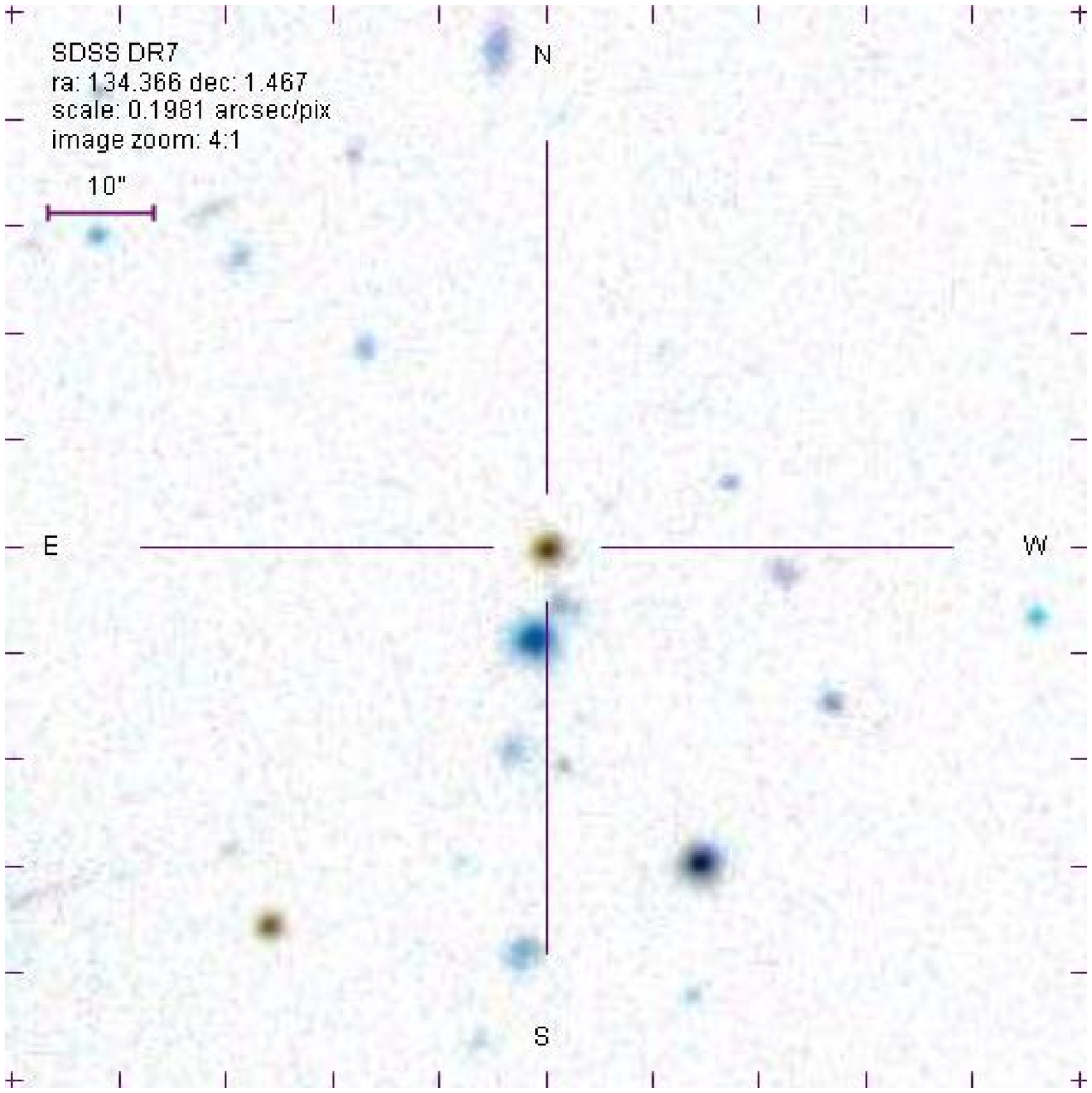}
   \includegraphics[width=4.7cm, angle=0]{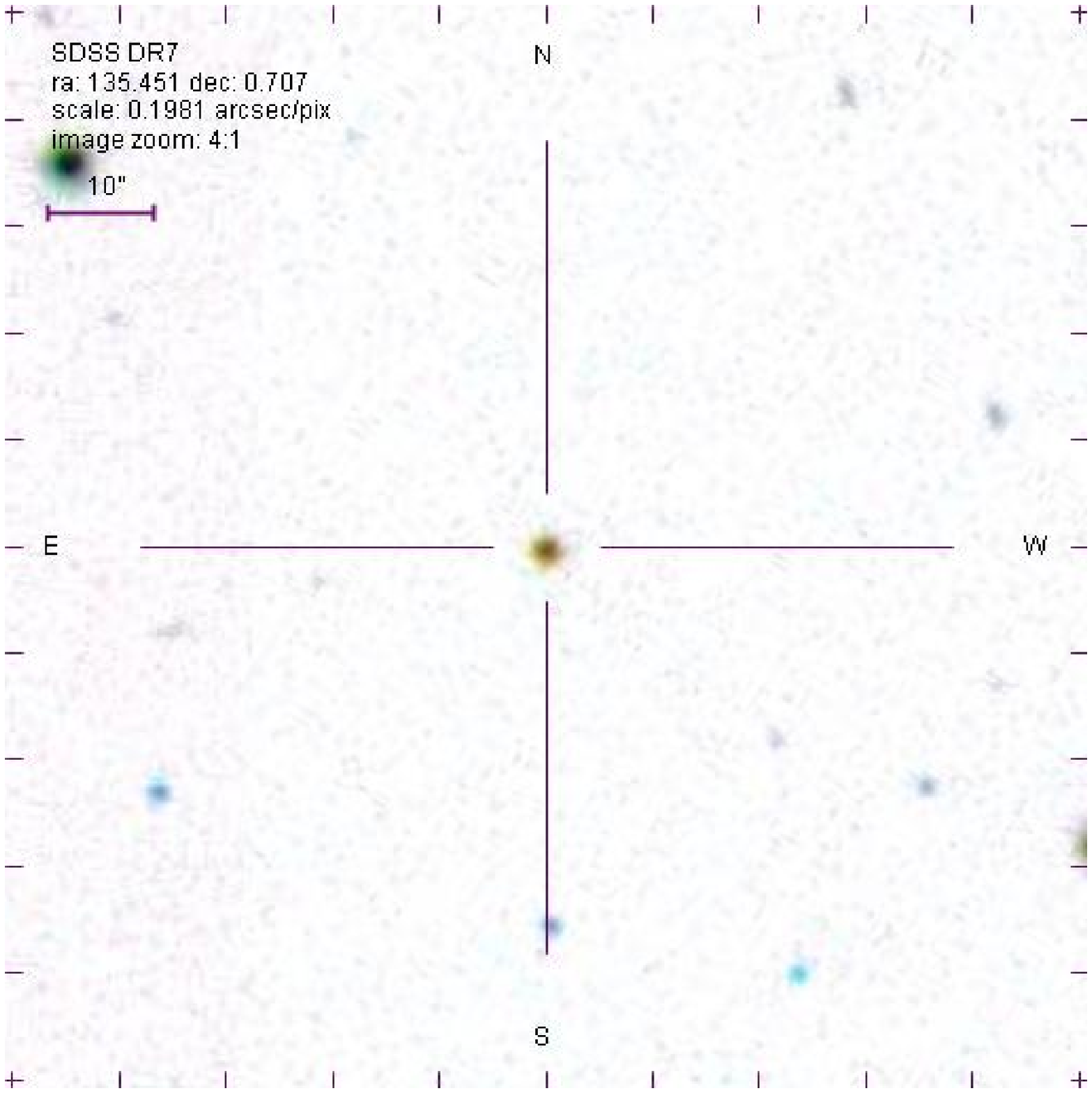}
   \includegraphics[width=4.7cm, angle=0]{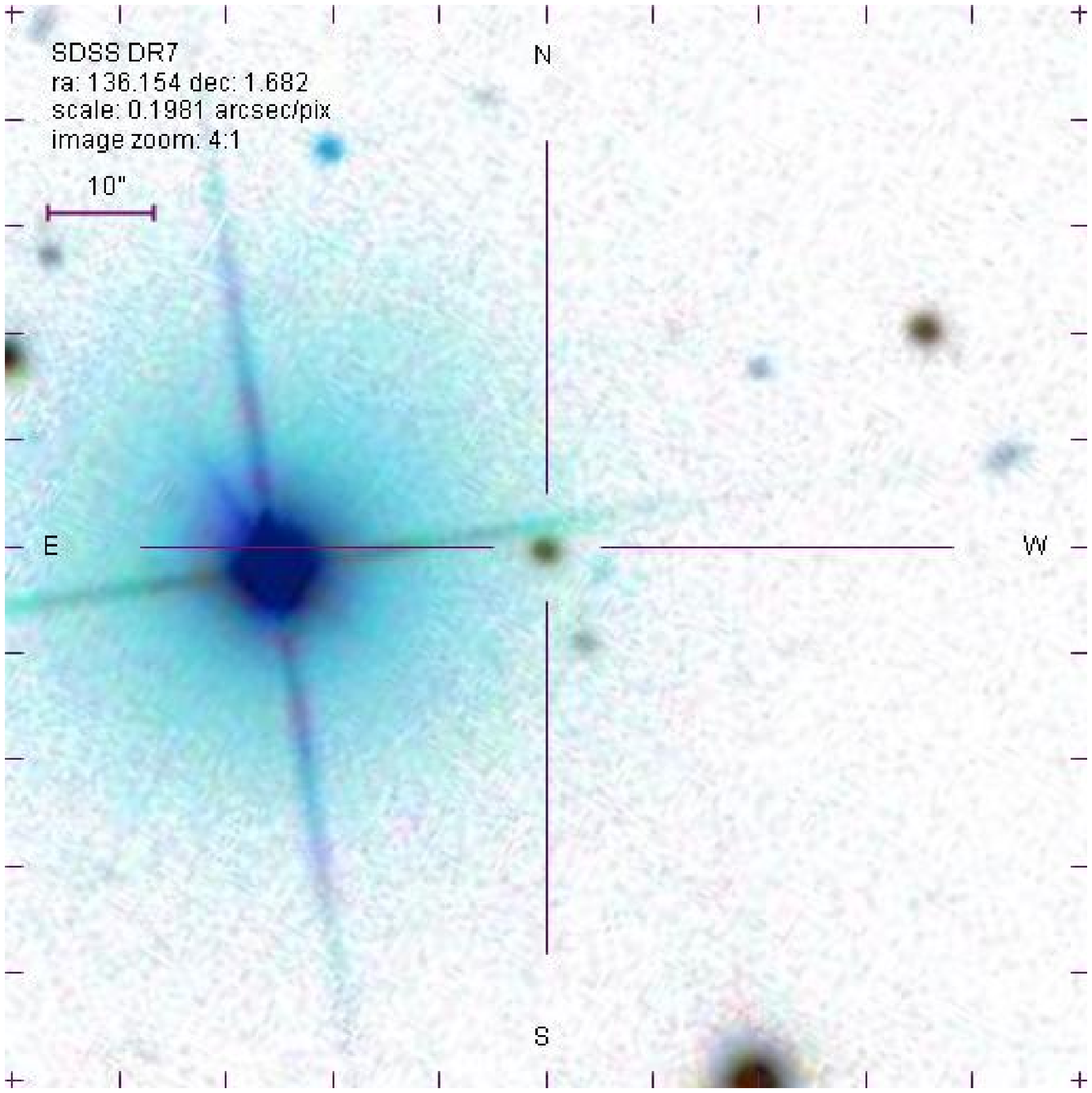}
    \includegraphics[width=4.7cm, angle=0]{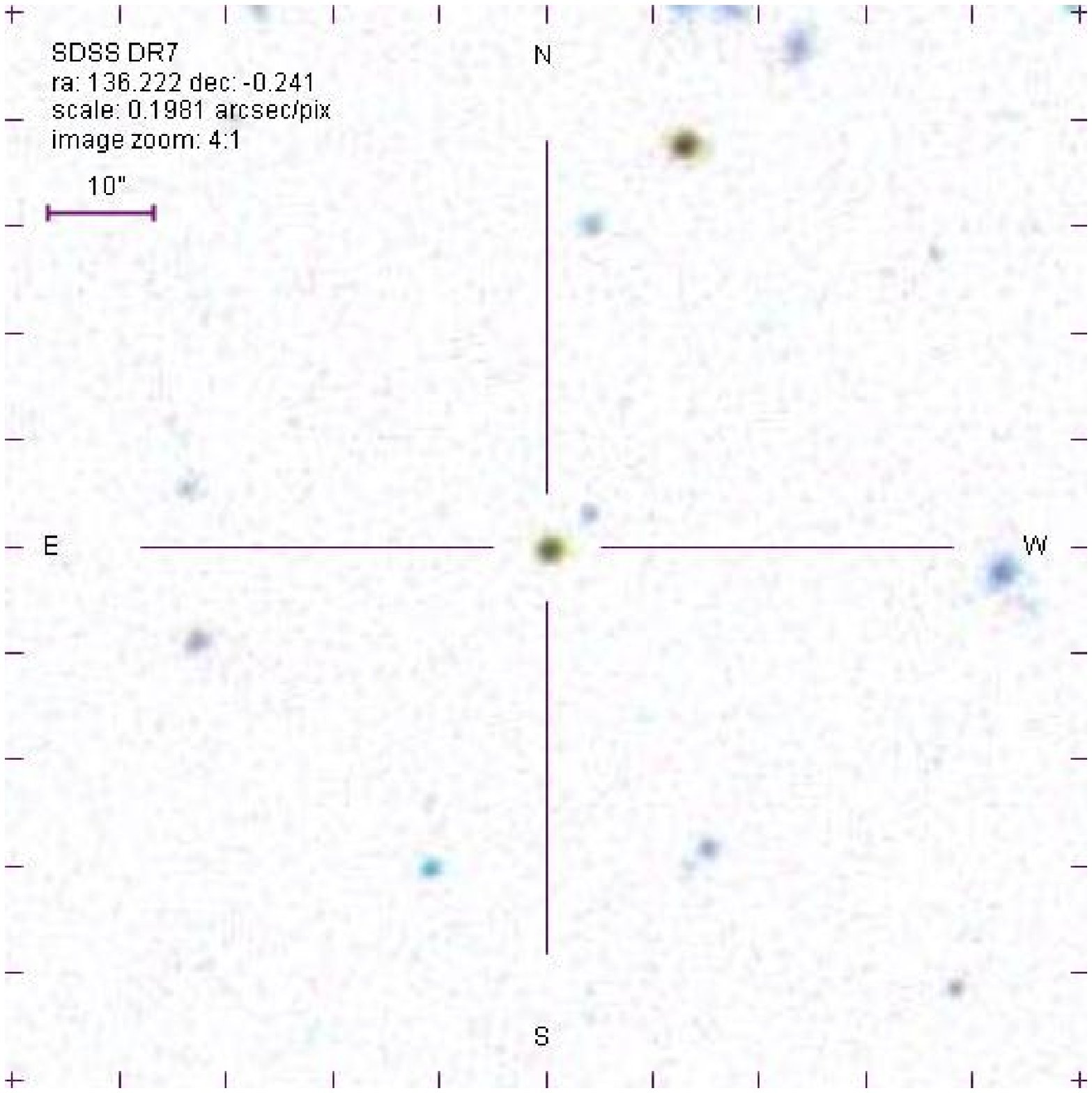}
    \includegraphics[width=4.7cm, angle=0]{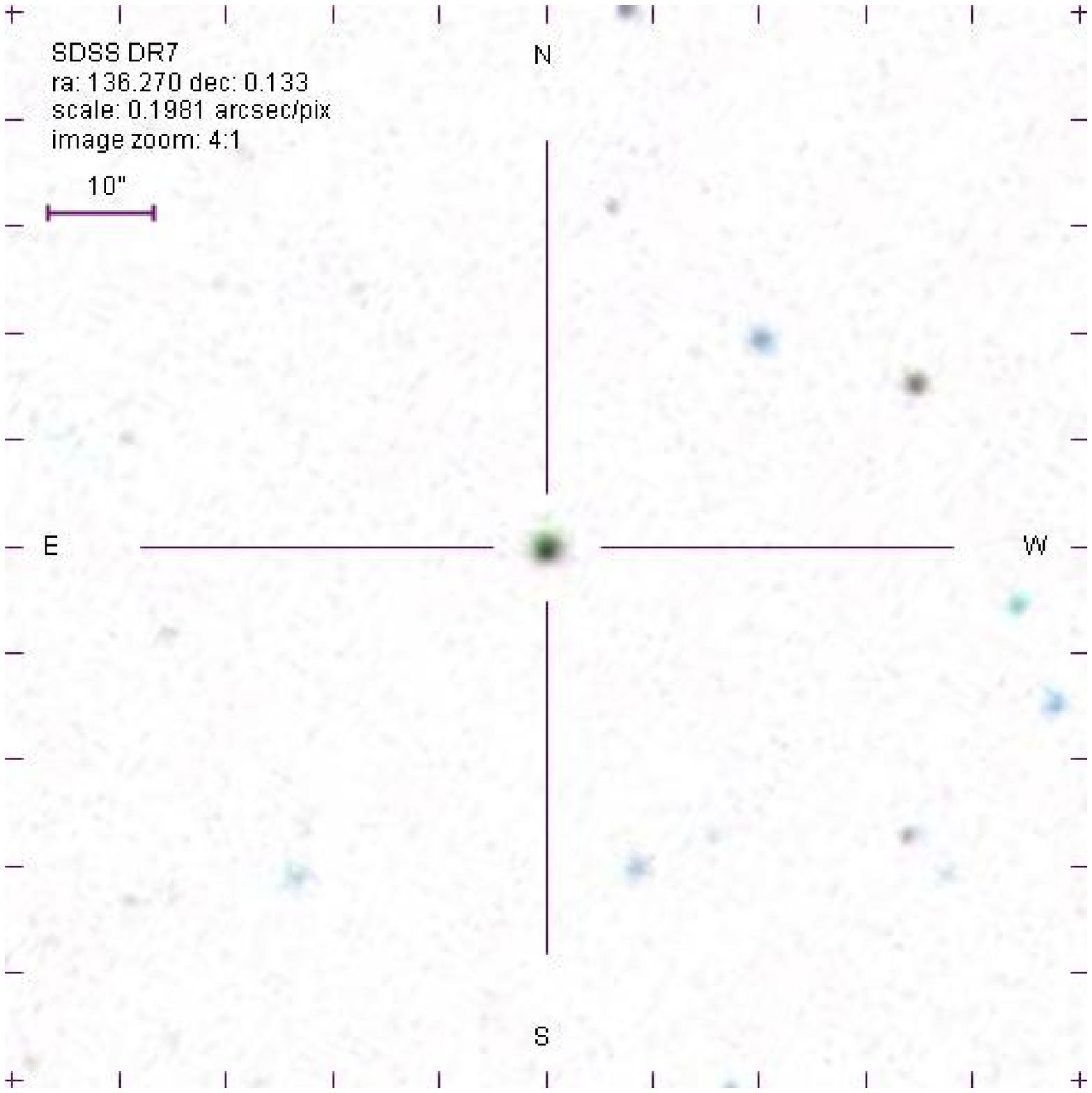}
   \caption{The finding charts of 8 new quasars are shown in an order of
increasing RA. The size of each chart is 100"$\times$100". } 
   \label{Fig1}
   \end{figure}

In Fig. 2 we show the LAMOST spectra of eight new quasars in an
order of increasing redshift (some sky light emissions
were not well subtracted).
The complicated feature around 5900$\AA$ in each spectrum is due to the 
problem in combining the LAMOST blue and red spectra, which overlap with 
each other from 5700$\AA$ to 6100$\AA$. 
From the spectra of six quasars with $z>1.3$, we can clearly identify
at least two broad emission lines and derive the average redshift
for them. For two quasars with $z<1.3$, only one emission line can be
reliably observed and is identified as MgII$\lambda 2798$. From the spectrum 
of SDSS J085718.29+024017.7 (z=1.154), we can actually see a line appeared 
around the wavelength of 4100$\AA$ although the S/N is not good in the blue
part. This is obviously the CIII]$\lambda 1909$ line and supports our
identification of the MgII line in the red part. Another support of these
identifications is from the photometric redshift estimation. For four of these
eight quasars, Richards et al. (2004) have given the photometric redshifts as
0.875, 1.075, 1.225 and 1.975, which is consistent with our spectroscopic
redshifts of 0.898, 1.154, 1.363 and 1.952, respectively.

   \begin{figure}
   \centering
   \includegraphics[width=15cm, height=16cm,angle=0]{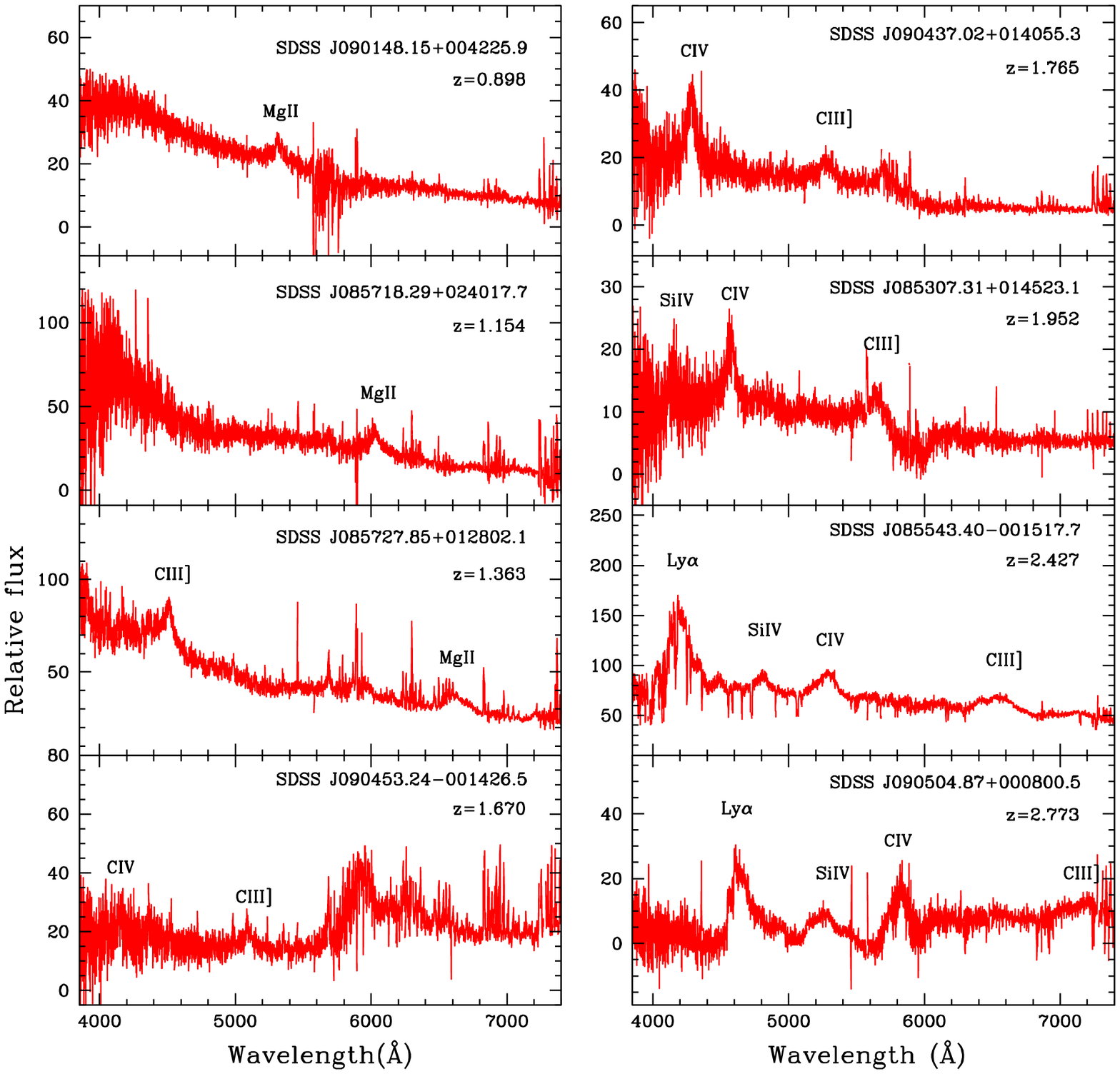}
   \caption{The LAMOST spectra of eight new quasars are shown in an order of
increasing reshift. The most prominent emission lines are marked in each 
spectrum.} 
   \label{Fig2}
   \end{figure}

 \begin{figure}
   \centering
   \includegraphics[width=15cm, angle=0]{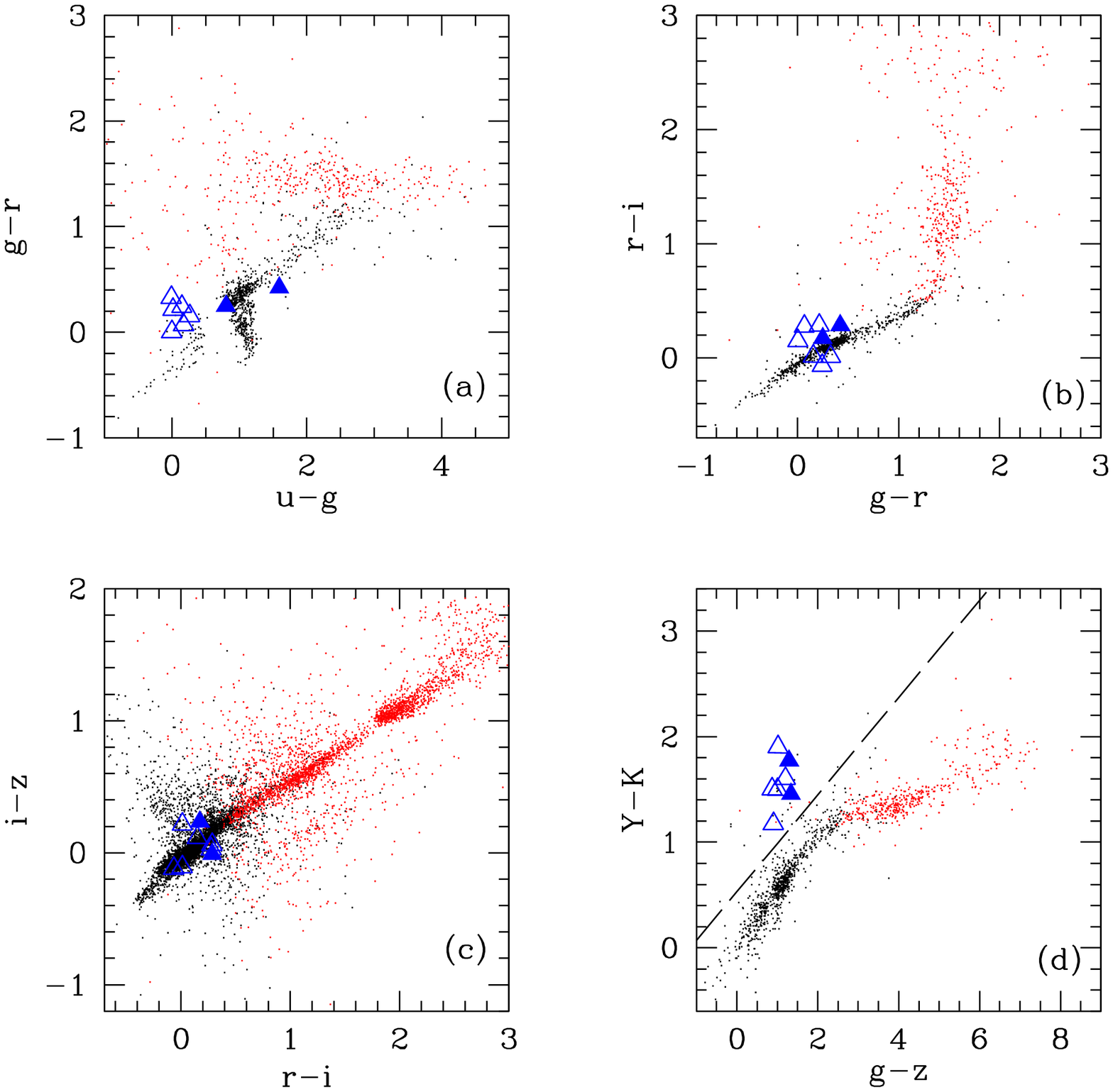}
   \caption{The location of two new $z>2.2$ quasars (solid triangles) and six new $z<2.2$ quasars
 (open triangles) in three optical color-color diagrams (a,b,c) and
the $Y-K$ vs. $g-z$ diagram (d), in comparing with the 8996 SDSS-UKIDSS stars (Wu \& Jia 2010). 
Black and red dots represent
the normal and later type stars, respectively. Dashed line shows the $z<4$ quasar selection criterion
proposed by Wu \& Jia (2010). In diagram (d) only seven quasars are shown because one quasar does not
have $Y$ band data.} 
   \label{Fig3}
   \end{figure}

We noticed that two of eight new quasars have redshifts larger than 2.2. 
Besides SDSS J085543.40-001517.7 ($z=2.427$) (see Paper I),  
SDSS J090504.87+000800.5 ($z=2.773$) is also a quasar in the 'redshift desert'. 
These quasars are very difficult
to be identified because of their similar optical colors as stars.
However, they can be recovered by using the near-IR colors. 
In Fig. 3 we show the locations of these eight quasars in three optical 
color-color diagrams and the $Y-K$ vs. $g-z$ diagram, in 
comparison with the 8996 SDSS-UKIDSS stars (Wu \& Jia 2010). 
Note that in the $Y-K$ vs. $g-z$ diagram the magnitude of $g$ and $z$ have been
converted to the magnitudes in Vega system by using the scalings 
(Hewett et al. 2006): 
$g=g(AB)+0.103$ and $z=z(AB)-0.533$. 
Obviously two quasars with redshifts larger than 2.2
locate in the stellar locus in all three optical color-color diagrams,
but are separated from stars in the  $Y-K$ vs. $g-z$ diagram and meets the 
selection criterion proposed by Wu \& Jia (2010). 
For six quasars with redshifts less than 2.2, although they are separated
from the main stellar locus in the $u-g$ vs. $g-r$ diagram, they still locate
in or close to the stellar locus in other two optical color-color diagrams.
None of these 8 new quasars has the SDSS spectrum, though 6 of them with 
redshifts less than 2.5 were
classified as quasar targets in the item of 'PrimeTarget' of the SDSS/DR7 database. 
These unidentified quasars in the SDSS spectroscopic survey
can be successfully recovered by applying the selection criterion in the
$Y-K$ vs. $g-z$ diagram.

We also searched the counterparts of these new quasars in other wavelength 
bands. From the VLA/FIRST radio catalog (White et al. 1997) we did not find 
radio counterparts for all eight quasars within 30$''$ from their
SDSS positions.  Therefore, these quasar are 
radio-quiet ones, which is another reason why they are not identified by the 
SDSS spectroscopy. 
We also searched the ROSAT X-ray source catalog (Voges et al. 1999) 
and did not find counterparts for them within 1'. From GALEX catalog 
(Morrissey et al, 2007) we 
found ultraviolet counterparts within 1$''$ from their
SDSS positions for five of six quasars with $z<2$. But for a  $z=1.363$
quasar, SDSS J085727.85+012802.1, and two quasars with $z>2.2$, we failed
to find their GALEX counterparts. The high GALEX detection rate (83\%) of $z<2$ 
quasars and the non-detection in ultraviolet for $z>2.2$ quasars in our case
is well consistent with 
the previous result of the SDSS-GALEX quasar sample (Trammell et al. 2007).

Although LAMOST met some problems during the commissioning 
observations, we were still able to identify other 38 known SDSS 
quasars in this field, with $i$ magnitudes from 16.24 to 19.10 and redshifts 
from 0.297 to 4.512.  The number of known SDSS quasars with $i<19.1$ 
in this five degree field is 177, and our identified 38 SDSS quasars take
a fraction of 22\% of them.  In the upper and lower panels of Fig. 4 
we show the histograms of redshift
distribution of 177 known SDSS quasars with $i<19.1$ and 38 SDSS quasars identified
by LAMOST in this field. The contributions of 8 new quasars to these two histograms 
are also demonstrated.
The ratio between 8 new quasars and 
38 known SDSS quasars identified by us
in this field is 21\%, impling that a substantial fraction of the quasars may be missed by
the SDSS at the magnitude limit $i<19.1$. 
Especially, only 4 of 177 SDSS known quasars with $i<19.1$ in this field have redshift 
larger than 2.4. Our discovery of 2 new quasars with $z>2.4$ adds a significant fraction 
of them.
Obviously this still needs to be confirmed
by more complete spectroscopic identifications of quasars in this field. In addition,
from the lower panel of Fig. 4 we can see that the fraction of quasars with redshifts around 1.2 is 
relatively lower, which is partly due to the lower CCD efficiency around $6000\AA$ where
the blue and red spectra overlap. If we take the spectrum of a quasar with $z\sim1.2$
with LAMOST, the MgII$\lambda 2798$ line will appear around $6000\AA$ as the only one prominent emission line
in the optical band but will be difficult to be identified
due to the current problems in combining the LAMOST blue and red spectra around $6000\AA$. This
situation will be improved after we solve the spectral combining problems.
 \begin{figure}
   \centering
   \includegraphics[width=13cm,angle=0]{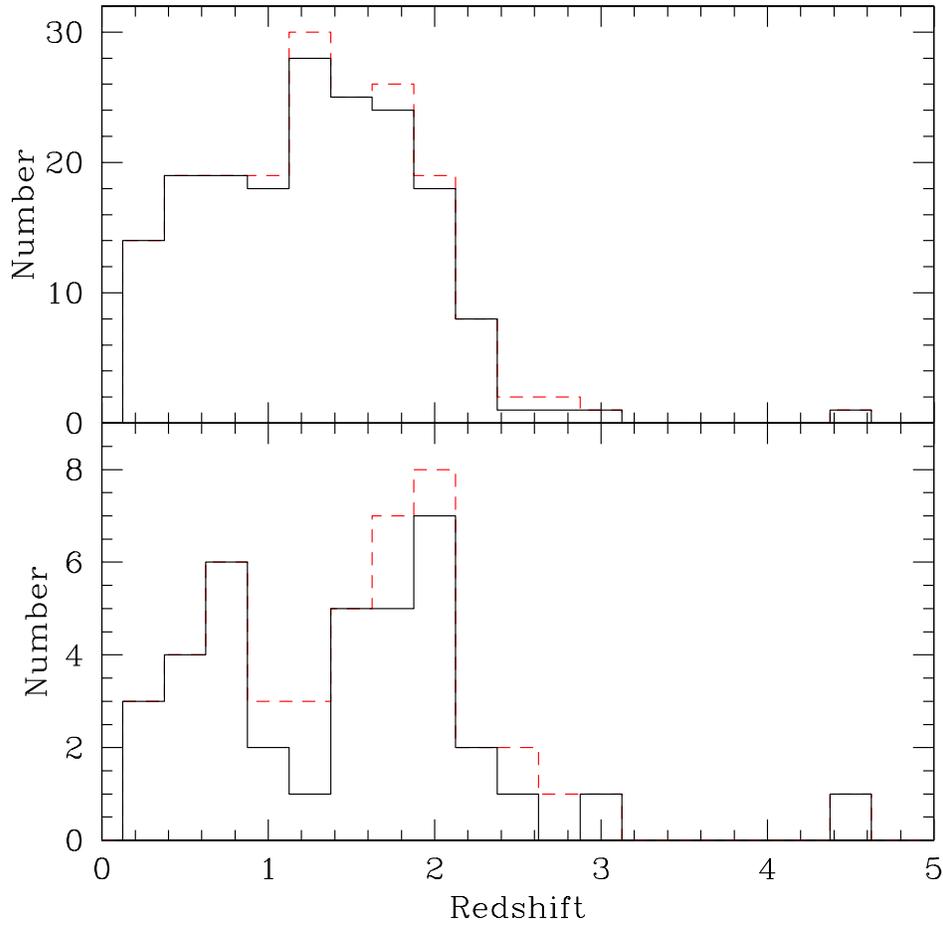}
   \caption{Upper panel: The histogram of redshift
distribution of 177 known SDSS quasars with $i<19.1$ in this field. Lower panel: 
The histogram of redshift
distribution of 38 SDSS quasars identified by LAMOST in this field. 
The contributions of 8 new quasars to these two histograms 
are also demonstrated by the dashed lines.} 
   \label{Fig4}
   \end{figure}

\section{Discussion}
\label{sect:discussion}
In this paper we presented the discovery of eight new quasars with redshifts
from 0.898 to 2.773 in one
extragalactic field close to GAMA-09 by the LAMOST commissioning observations. 
This discovery supports the idea that by combining the UKIDSS near-IR colors with the
SDSS optical colors we are able to efficiently recover the unindentified quasars in
the SDSS spectroscopic survey even at the magntude limit $i<19.1$.
Our results indicate that not only some quasars in the redshift desert 
but also some quasars with lower redshifts are probably missed in the SDSS survey.
These missing quasars may take a substantial fraction of the quasars at
the magnitude limit of SDSS spectroscopy. Obviously this still needs to be confirmed
by more complete identifications of quasars in this field, because
our identifications during the LAMOST commissioning observations are incomplete.

Nevertheless, the success of identifing eight new quasars (including two quasars
in the redshift desert) in one
extragalactic field gives us more confidence to discover more missing quasars 
in the future LAMOST observations. In the winter of 2009, LAMOST has made test
observations on several sky fields and we are now searching for more quasars from
the spectra taken in these fields. We believe that more missing quasars 
will be discovered soon. 

A complete quasar sample is very important to the construction of the quasar 
luminosity function and study the cosmological evolution of quasars.
However, as we demonstrated in this paper, because some quasars have similar 
optical colors as normal stars, it is very difficult for find them in the optical 
quasar surveys. The low efficiency of finding quasars in the redshift desert 
($z$ from 2.2 to 3) has led to obvious incompleteness of SDSS quasar sample 
in this redshift range and serious problems in 
constructing the luminosity function for quasars around the redshift peak 
(between 2 and 3) of quasar activity (Richards et al. 2006; Jiang et al. 2006). 
Therefore, recovering
these missing quasars will become an important task in the future quasar survey.
We hope that in the next a few months great
progress will be made in improving the capability of LAMOST spectroscopy. 
As long as LAMOST 
can reach its designed capability after the commissioning phase, we expect to obtain
the largest quasar sample in the LAMOST quasar survey. This sample will 
undoubtedly play a leading role in the future quasar study.

\normalem
\begin{acknowledgements}
We thank Michael Strauss for clarifying the SDSS target status of these new quasars. 
This work was supported by the National 
Natural Science Foundation of China  (10525313), the
National Key Basic Research Science Foundation of China (2007CB815405).
The Large Sky Area Multi-Object Fiber Spectroscopic Telescope (LAMOST)
is a National Major Scientific Project built by the Chinese Academy of
Sciences. Funding for the project has been provided by the National
Development and Reform Commission. The LAMOST is operated and managed
by the National Astronomical Observatories, Chinese Academy of Sciences.
We acknowledge the use of LAMOST,  as well as
the archive data from SDSS, UKIDSS, FIRST, ROSAT and GALEX. 

\end{acknowledgements}

\end{document}